\documentclass[prl,twocolumn,showpacs,preprintnumbers,amsmath,amssymb]{revtex4}
\usepackage{graphicx}
\usepackage{dcolumn}
\usepackage{bm}

\begin{document}

\title{Scalable Superconducting Architecture for Adiabatic Quantum Computation}

\author{William M. Kaminsky}
 \email{wmk@mit.edu}
 \affiliation{Department of Physics, Massachusetts Institute of Technology, Cambridge, Massachusetts 02139}
\author{Seth Lloyd}
 \affiliation{Department of Mechanical Engineering, Massachusetts Institute of Technology, Cambridge, Massachusetts 02139}%
\author{Terry P. Orlando}
 \affiliation{Department of Electrical Engineering and Computer Science, Massachusetts Institute of Technology, Cambridge, Massachusetts 02139 }

\date{\today}

\begin{abstract}
A scalable superconducting architecture for adiabatic quantum
computers is proposed.  The architecture is based on
time-independent, nearest-neighbor interqubit couplings: it can
handle any problem in the class NP even in the presence of
measurement errors, noise, and decoherence. The implementation of
this architecture with superconducting persistent-current qubits
and the natural robustness of such an implementation to
manufacturing imprecision and decoherence are discussed.
\end{abstract}


\maketitle


Adiabatic quantum computation \cite{Farhi} is an approach to
solving computational problems of the complexity class NP
\cite{Garey} via energy minimization. In particular, by exploiting
the ability of coherent quantum systems to follow adiabatically
the ground state of a slowly changing Hamiltonian, it aims to
bypass the many separated local minima that occur in difficult
minimization problems. Adiabatic quantum computation is of
theoretical interest because it provides a straightforward,
non-oracular way to pose class NP problems on a quantum computer.
To date, most research on it focuses on ascertaining its time
complexity \cite{Farhi,ChildsQIC,Hogg,Vazirani,Farhi2}. However,
it is also of practical interest because encoding a quantum
computation in a single eigenstate, the ground state, offers
intrinsic protection against dephasing and dissipation
\cite{ChildsPRA,RolandCerf}.

In this Letter, we present a scalable superconducting architecture
for adiabatic quantum computation that can handle any class NP
problem. It requires neither efficient qubit measurements, nor
interqubit couplings beyond nearest neighbors, nor couplings that
vary during the course of the computation. We also discuss how to
implement this architecture specifically with superconducting
persistent-current (PC) qubits \cite{Mooij,Orlando}, which
constitute a promising approach to lithographable solid-state
qubits. We show that the proposed architecture is robust against
manufacturing imprecision, and we estimate the maximum size
problem the architecture could support if it were implemented with
existing PC qubits at dilution refrigerator temperatures of 10 mK.
This maximum stems from the condition that the environment must
not excite the computer from its ground state.  A simple Boltzmann
factor argument implies that a 10 mK temperature limits the PC
qubit architecture to NP problem instances of $O(50)$ bits.
However, advances in cryogenics or in fabricating PC qubits from
higher-$T_c$ materials could raise this limit by 1 or 2 orders of
magnitude.


\textit{Background}: In adiabatic quantum computation, one encodes
the answer(s) to a hard constrained minimization problem in the
ground state(s) of a suitable Hamiltonian $H$ whose local
couplings ensure its ground state(s) satisfy the problem's
constraints. One initiates the system in the ground state of
another Hamiltonian $H_0$ chosen so its ground state is quickly
reachable simply by cooling.  One then adiabatically deforms $H_0$
into $H$ by applying a time-dependent Hamiltonian $H(t)$ such that
$H(0) = H_0$ and $H(T) = H$.  The adiabatic approximation holds as
long $H(t)$ possesses at all times a spectral gap $\Delta(t)$
between its instantaneous ground $|\psi_0(t)\rangle$ and excited
states $|\psi_n(t)\rangle$ such that $\frac{\langle \psi_0(t) |
dH^2(t)/dt | \psi_0(t) \rangle}{\Delta^2(t)} \ll 1$. The question
of whether adiabatic quantum computation has polynomial or
exponential time complexity is thus determined by whether the
minimum gap $\Delta_{\min}(n)$ shrinks polynomially or
exponentially in the number of qubits $n$.

It is still unknown what speedup adiabatic quantum computation
offers in general over classical energy minimization algorithms.
Numerical investigations of hard instances of NP-complete problems
\cite{Farhi,ChildsQIC,Hogg} suggest that $\Delta_{\min}(n) =
O(n^{-1})$, at least typically, and thus adiabatic quantum
computation may provide for all practical purposes an exponential
speedup on NP-complete problems over all known classical
algorithms.  However, quantitatively bounding the minimum gaps for
adiabatic algorithms for NP-complete problem instances is
apparently as hard as solving the actual NP-complete instances
themselves. Thus, these numerical investigations have been
confined to problems with $\alt 30$ qubits. Ref.~\cite{Vazirani}
constructs a problem that takes exponential time for a simple
version of the adiabatic algorithm. However, Ref.~\cite{Farhi2}
shows that alternative versions of the adiabatic algorithm solve
the problem of \cite{Vazirani} in polynomial time, and presents a
problem for which simulated annealing provably takes exponential
time yet adiabatic quantum computation takes only polynomial time.
Definitively establishing the efficiency of the adiabatic
algorithm for large problem instances may thus have to await the
actual construction of adiabatic quantum computers of the sort
described here.


\textit{Layout Requirements:} By definition, a scalable
programmable architecture for adiabatic quantum computing
possesses some regular geometrical layout of switchable pairwise
couplings on $m$ nodes that can efficiently encode problem
Hamiltonians corresponding to any instance of some desired class
of energy minimization problems up to $n \leq m$ bits in size. To
have the maximum flexibility in the problems one can pose and to
exploit the maximum potential power of adiabatic quantum
computation, the binary constraint problem the architecture
naturally poses should be NP-complete, and the architecture should
readily pose instances of it that are truly difficult for all
known classical heuristics. Additionally, present experimental
constraints make it highly desirable to have couplings that
neither extend beyond nearest neighbors nor vary in time beyond
possibly switching between a full strength ``on'' state and a much
reduced strength ``off'' state before computation begins so as to
allow programming of the desired problem.

An architecture meeting these requirements naturally follows from
the fact \cite{Barahona} that calculating the ground state of an
antiferromagnetically coupled Ising model in a uniform magnetic
field is isomorphic to solving the NP-complete graph theory
problem Max Independent Set (MIS), which is the problem of finding
for a graph $G = (V,E)$ the largest subset $S$ of the vertices $V$
such that no two members of $S$ are joined by an edge from $E$.
Interqubit couplings may be laid on a single plane and kept to a
modest number since MIS remains NP-complete even in the ostensibly
simple, topologically uniform case of degree-3 planar graphs,
\textit{i.e.}, graphs that can be drawn in a plane without any
edges crossing and in which every vertex is connected to exactly 3
others \cite{Garey2}.

The isomorphism takes a particularly simple form in this case: the
MISs of a degree-3 planar graph $G = (V,E)$ are the ground states
of an Ising model with spins on vertices $V$, equal strength
antiferromagnetic couplings along some desired axis $\hat{n}$ on
edges $E$, placed in a uniform applied magnetic field also along
$\hat{n}$:
\begin{equation}
H \propto \sum_{i \in V}\sigma_{\hat{n}}^{(i)} + \sum_{i,j \in
E}\sigma_{\hat{n}}^{(i)}\sigma_{\hat{n}}^{(j)}
\end{equation}

In regard to the requirement of posing truly hard problem
instances, note that the most efficient classical approximation
algorithm known for MIS restricted to planar graphs has a cost
that grows exponentially in the desired accuracy. Specifically,
the cost to obtain an approximate MIS of a planar graph with $n$
vertices that has a cardinality which is at least $\frac{k}{k+1}$
of the true MIS's cardinality is $O(8^k k n)$, and is thus
impractical once one desires $\agt 90\%$ accuracy ($k \agt 9$)
\cite{Baker}.

The requirements of nearest neighbor couplings, maximum
uniformity, and time-independent control can be met via the layout
depicted in Fig.~1: a triangular lattice with qubits at its
vertices and nearest-neighbor ferro- or antiferromagnetic
couplings on its edges. To allow programmability, that is, to
embed an arbitrary degree-3 planar graph in this triangular
lattice, each coupling is switchable and moreover the magnetic
field on each qubit is individually controllable so as to allow
one to create ``dummy'' qubits that possess no single qubit
Hamiltonian and act simply to propagate a coupling. (Triangular
lattices make it especially easy to embed degree-3 planar graphs,
but lattices with a lower node degree, such as square or hexagonal
lattices, would also suffice.)

\begin{figure}
\includegraphics[width=3.4in,height=2.5in]{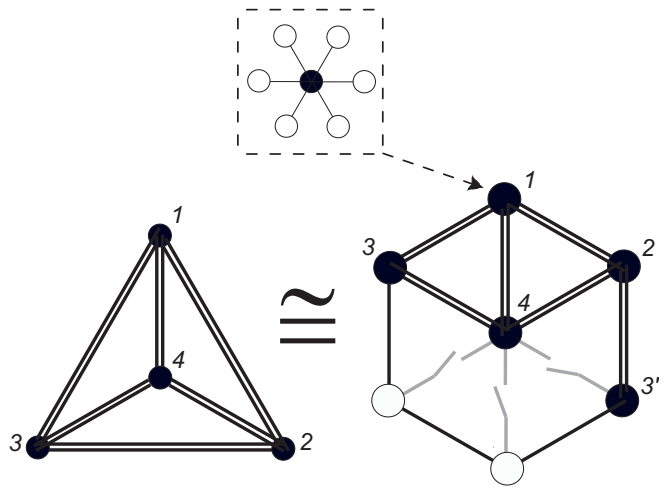}
\caption{Schematic of the embedding of the simplest degree-3
planar graph (\textit{left}) into the architecture
(\textit{right}). The architecture is built on a triangular
lattice with qubits at its vertices (schematically denoted by
filled circles if possessing a Zeeman term and thus denoting a
graph vertex, open circles if not and thus denoting a dummy vertex
that serves only to propagate a coupling) and switchable
nearest-neighbor couplings on its edges (schematically denoted by
black double lines if antiferromagnetic and switched on, black
single lines if ferromagnetic and switched on, and gray broken
lines if switched off). This allows an arbitrary degree-3 planar
graph to be embedded efficiently within the lattice, as can be
seen in how the Qubit $\#3$ of the graph is ferromagnetically
coupled via dummies to a partner $\#3'$ so that is a nearest
neighbor of $\#2$. \textit{Inset}: To compensate for measurement
errors, the logical qubits in the architecture (denoted by the
larger circles) are actually ferromagnetically coupled clusters of
physical qubits (smaller circles).   For example, as depicted
above, one could ferromagnetically couple 6 dummies to each
physical computational qubit and thus essentially compensate for
measurement errors by a classical 7-bit repetition code.  The
ferromagnetic couplings in such redundant clusters need not be
switchable.}
\end{figure}


\textit{Implementation with PC Qubits}: Fig.~2A depicts the
specific PC qubit circuit we shall consider here.
Ref.~\cite{Orlando} explains the rationale underlying its design
and its canonical quantization. For our purposes, the pertinent
result of Ref.~\cite{Orlando} is that in the regime of
sufficiently low temperatures ($T \alt 10^{-3} E_J \approx 30$ mK
for our parameters) and frustrations that make all the minima of
the junctions' potential energies nearly equal ($0.485 \Phi_0 \alt
f^{\mathrm{bot}} + \frac{1}{2}f^{\mathrm{top}} \alt 0.515
\Phi_0)$, the circuit is accurately described by a truncated
2-level qubit Hamiltonian that has a simple, readily tunable form.
Specifically, numerical analysis shows for realistic parameters of
$E_J / E_C = 80, \beta = 0.8, \gamma = 0.02$ and frustrated at
$f^{\mathrm{top(bot)}} = \left[ 0.330 + \delta^{\mathrm{top(bot)}}
\right] \Phi_0$ yields an effective qubit Hamiltonian $H_Q$:
\begin{equation}
\frac{H_Q \left(\delta^{\mathrm{top}},\delta^{\mathrm{bot}}
\right)}{E_J} = K_z
\left(\delta^{\mathrm{top}},\delta^{\mathrm{bot}} \right) \sigma_z
+ K_x \left(\delta^{\mathrm{top}},\delta^{\mathrm{bot}} \right)
\sigma_x
\end{equation}
where
\begin{eqnarray}
K_z \left(\delta^{\mathrm{top}},\delta^{\mathrm{bot}} \right) &=&
-0.025 + 3.8 \delta^{\mathrm{bot}} + 2.0
\delta^{\mathrm{top}} \\
K_x \left(\delta^{\mathrm{top}},\delta^{\mathrm{bot}} \right) &=&
0.0049 - 1.2 \delta^{\mathrm{bot}} - 0.81 \delta^{\mathrm{top}}
\end{eqnarray}
and the ``z-basis'' is the basis of classical counter-rotating
current states, $\left\{ \mid \circlearrowleft \rangle, \mid
\circlearrowright \rangle \right\}$, while the ``x-basis'' is the
basis of real symmetric and antisymmetric linear combinations of
the classical states, $\left\{\mid \circlearrowleft \rangle + \mid
\circlearrowright \rangle, \mid \circlearrowleft \rangle - \mid
\circlearrowright \rangle \right\}$.

\begin{figure}
\includegraphics[height=2.5in]{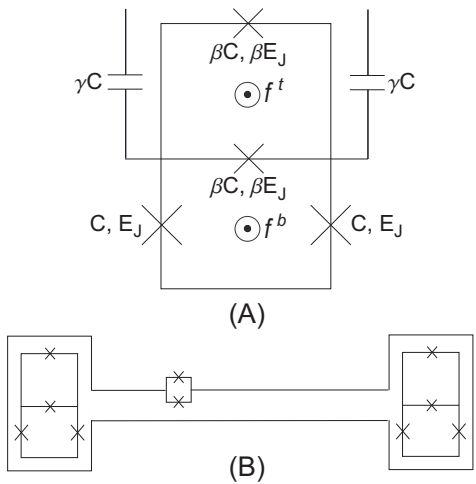}
\caption{(A) Circuit schematic for PC qubit, after
Ref.~\cite{Orlando}.  (B) Schematic (not to scale) of two qubits
coupled by inductive coils connected by a bus possessing a switch
such as a SQUID (depicted) \cite{Mooij} or a JoFET \cite{Storcz}.
Qubits are separated so that free space inductive coupling is
negligible, and coupling via the inductive coils dominates.  If
the coils trap zero net flux, if $M_{TQ}$ denotes the mutual
inductance between the flux transformer and a loop of the qubit,
and if $L_T$ denotes the self inductance of the flux transformer,
then the effective mutual inductance between the qubits $M_{QQ}
\approx M_{TQ}^2 / L_T$ and the effective qubit self inductance
$L_Q$ is largely suppressed.}
\end{figure}

Beyond this regime where Equ.~(5) is valid (\textit{i.e.},
$|\delta^{\mathrm{top}} + \frac{1}{2} \delta^{\mathrm{bot}}| \agt
0.015$), the circuit overwhelmingly favors a single circulation
direction for its current, meaning in the reduced 2-level picture
that the $\sigma_z$ component dominates over the $\sigma_x$ by
well over a factor of 20.  Such a $\sigma_z$ component can
dominate over all couplings, naturally providing a useable
starting Hamiltonian with a ground state that is reachable by
simple cooling despite the presence of couplings that are always
switched on.

Qubits are coupled inductively as in Fig.~2B.   Switching of
couplings between a full strength ``on'' and a much reduced
strength ``off'' could be accomplished, for example, magnetically
via DC SQUIDs \cite{Mooij} or electrostatically via JoFETs
(Josephson Field-Effect Transistors) \cite{Storcz}.  Now consider
an inductively coupled pair of identical PC qubits with identical
applied frustration offsets $\delta^{\mathrm{top}}_{1(2)} \equiv
\delta^{\mathrm{top}},\delta^{\mathrm{bot}}_{1(2)} \equiv
\delta^{\mathrm{bot}}$. Moreover, for simplicity, let the mutual
inductances from each qubit to either loop of the other be
identical, $L^{\mathrm{top(bot)}}_{12(21)} \equiv M$.  Inductive
effects are calculated perturbatively to lowest order by equating
the qubit circulating current to the mean current travelling
through the DC SQUID in the circuit, which the previously cited
single qubit numerical simulations show is roughly $\pm 1.4 I_c
\cos[\pi(0.330 + \delta^t)]$.

Within this approximation, the basic building block of our desired
problem Hamiltonian of a Max-Independent-Set-encoding
antiferromagnetic Ising model in a uniform field, Equ.~(3), is
achieved at an operating point $\left(\delta^t_C =
-0.0124,\delta^b_C = 0.0200 \right)$ with $E_J/(MI_c^2) = 90$:
\begin{equation}
\frac{H_C}{E_J} = 0.029 \left( \sigma_{\hat{c}1} +
\sigma_{\hat{c}2} + \sigma_{\hat{c}1} \sigma_{\hat{c}2} \right)
\end{equation}
where $\hat{c}$ is the vector rotated $19^\circ$ clockwise from
the $z$-axis in the $xz$-plane of the Bloch sphere.

Realistic parameters for a Nb PC qubit are $E_J = 0.60$ THz and
$I_c = 1.2 \mu$A.  Therefore, the desired ratio $E_J/(MI_c^2) =
90$ requires a mutual inductance of $M = 3.1$ pH, which is also
realistic.

We now turn to the problem of constructing dummy qubits that only
propagate ferromagnetic couplings and do not possess single qubit
terms.  Ideally, these ferromagnetic couplings would be
proportional to $-\sigma_{\hat{c}1} \sigma_{\hat{c}2}$ and thus
commute with $H_C$. However, if we constrain our inductive
couplings to have only 2 settings: ``on'' where $M = 3.1$ pH and
``off'' where $M \approx 0$ pH, then the operating point
$\left(\delta^t_D = -0.0171,\delta^b_D = 0.0152 \right)$ that
causes the single qubit terms to vanish will not yield a coupling
$H_D \propto -\sigma_{\hat{c}1} \sigma_{\hat{c}2}$. However, this
poses no practical problem as the actual coupling is very close to
$-\sigma_{\hat{c}1} \sigma_{\hat{c}2}$:
\begin{equation}
\frac{H_D}{E_J} = -(3.6 \times 10^{-3})~\sigma_{\hat{d}1}
\sigma_{\hat{d}2} + (5.4 \times 10^{-5}) \sigma_{x1} \sigma_{x2}
\end{equation}
where $\hat{d}$ is the vector rotated $16^\circ$ clockwise
(\textit{i.e.,} $3^\circ$ less than $\hat{c}$) from the $z$-axis
in the $xz$-plane of the Bloch sphere.  (It probably unwise at
present to expend any effort toward making $H_D$ closer to
$-\sigma_{\hat{c}1} \sigma_{\hat{c}2}$ given the likely level of
accuracy of our formulas for the Hamiltonian.)

Similarly, the Hamiltonian $H_{CD}$ for a computational qubit
inductively coupled to a dummy qubit is not in the ideal form
$-\sigma_{\hat{c}1} \sigma_{\hat{d}2}$, but again it is tolerably
close:
\begin{equation}
\frac{H_{CD}}{E_J} = 0.029 \sigma_{\hat{c}1} - 0.013
\sigma_{\hat{c}1} \sigma_{\hat{e}2}
\end{equation}
where $\hat{e}$ is the vector rotated $24^\circ$ clockwise
(\textit{i.e.,} $8^\circ$ more than $\hat{d}$) from the $z$-axis
in the Bloch sphere's $xz$-plane.

The adiabatic computation is performed simply by slowly bringing
$\left(\delta^{\mathrm{top}}, \delta^{\mathrm{bot}} \right)$ on
all the computational qubits from any convenient point with
$\delta^{\mathrm{top}} + \frac{1}{2} \delta^{\mathrm{bot}} \agt
0.015$ to the operating point $(\delta^{\mathrm{top}}_C,
\delta^{\mathrm{bot}}_C)$ while keeping all dummy qubits fixed at
their operating point
$(\delta^{\mathrm{top}}_D,\delta^{\mathrm{bot}}_D)$.

The intrinsic robustness of adiabatic quantum computation versus
environmental noise has been demonstrated both numerically
\cite{ChildsPRA} and theoretically \cite{RolandCerf}. However,
measurement error is presently a critical concern with PC qubits
because their most conveniently measured characteristic is the
flux created by their persistent currents.  This limits them to be
measured in what we have called the $z$-basis despite the fact
that it often will be necessary to work it a different basis for
computation. Specifically in the case of this architecture, the
computational basis is rotated $19^\circ$ away from the $z$-axis.
Therefore, any measurement scheme for this architecture based on
measuring a qubit's magnetic flux will have an error probability
of at least $\sin^2(19^\circ) = 0.11$. Moreover, a special concern
arises in adiabatic computation based on frustrated Ising models
since such systems generically have highly degenerate ground
states. Simple repetition of the algorithm will thus generically
yield a different ground state with each measurement. Measurement
errors cannot be corrected by averaging such uncorrelated data
together. It is therefore necessary to program correlated
redundancy into the architecture by having dummy qubits
ferromagnetically coupled to each computational qubit. Thus, when
measurement collapses the computer's state, one will obtain
multiple copies of one valid solution, and then averaging the data
from measuring all these added dummy qubits will compensate for
measurement errors via a classical repetition code.   The
couplings between these redundant dummies and the computational
qubits need not be switchable, and can be designed in such a way
that the addition of redundancy does not decrease the minimum gap
$\Delta_{\min}$.

The key constraint on the number of qubits in an adiabatic quantum
computer is that the environment must not excite the computer out
of its ground state. Conservatively, this imposes a limit
$n_{\max}$ on the number of logical qubits such that
$\Delta_{\min}(n_{\max}) > kT$, the environment's average thermal
energy.  As the form of $\Delta_{\min}(n)$ is still an open
problem, no exact answer can be given presently. However, as cited
previously, there are indications that $\Delta_{\min}(n) =
O(n^{-1})$, at least typically \cite{Farhi,ChildsQIC,Hogg}. If
this is true asymptotically, then the maximum number of qubits
given an operating temperature $T$ is $O[\Delta_1 / (kT) ]$ where
$\Delta_1$ is the energy gap of the problem Hamiltonian for a
single computational qubit. For the previously cited parameters
for existing Nb PC qubits, $\Delta_1 = 18$ GHz. As dilution
refrigerators can bring the electron temperature to 15-20 mK, the
above assumptions imply $n_{\max} = O(50)$. Such a value
essentially is the maximum possible at dilution refrigerator
temperatures with Nb PC qubits of any possible design for it
saturates the limit on $\Delta_1$ set by the Nb's critical
temperature $\Delta_1 = O(0.1kT_c) = O(0.9 \hbox { K})$.  Advances
such as building Josephson junctions out of higher-$T_c$ materials
and/or achieving electronic temperatures in superconductors of
$O(1 \hbox{ mK})$ could increase this conservative estimate of
$n_{\max}$ to hundreds or thousands.

Finally, the architecture is robust versus manufacturing
imprecision for three reasons.  First, any undesired term in the
Hamiltonian that is a product of Pauli operators will couple a
problem's solution to only one other excited state.  Perturbation
theory therefore implies that for given energy scalar $\kappa$
[see Equ.~(5)], tolerance for inaccuracies in the couplings
decreases only linearly in $n$. Second, since all PC qubits have
two controllable parameters, $\delta^{\mathrm{top}}$ and
$\delta^{\mathrm{bot}}$, certain fabrication errors can be
compensated by individually calibrating $\delta^{\mathrm{top}}$
and $\delta^{\mathrm{bot}}$ for each qubit. Third, if some qubits
completely malfunction, the redundancy in the triangular grid
layout allows one to steer coupling chains around them.

\begin{acknowledgements}
This work is supported in part by the AFOSR/NM grant
F49620-01-1-0461 and in part by DARPA under the QuIST program. WMK
thanks the Fannie and John Hertz Foundation for its fellowship
support.
\end{acknowledgements}

\bibliography{WMK_SL_TPO_Adiabat_Arch_v2}

\begin{thebibliography}{14}
\expandafter\ifx\csname natexlab\endcsname\relax\def\natexlab#1{#1}\fi
\expandafter\ifx\csname bibnamefont\endcsname\relax
  \def\bibnamefont#1{#1}\fi
\expandafter\ifx\csname bibfnamefont\endcsname\relax
  \def\bibfnamefont#1{#1}\fi
\expandafter\ifx\csname citenamefont\endcsname\relax
  \def\citenamefont#1{#1}\fi
\expandafter\ifx\csname url\endcsname\relax
  \def\url#1{\texttt{#1}}\fi
\expandafter\ifx\csname urlprefix\endcsname\relax\def\urlprefix{URL }\fi
\providecommand{\bibinfo}[2]{#2}
\providecommand{\eprint}[2][]{\url{#2}}

\bibitem[{\citenamefont{Farhi et~al.}(2001)\citenamefont{Farhi, Goldstone,
  Gutmann, Lapan, Lundgren, and Preda}}]{Farhi}
\bibinfo{author}{\bibfnamefont{E.}~\bibnamefont{Farhi}},
  \bibinfo{author}{\bibfnamefont{J.}~\bibnamefont{Goldstone}},
  \bibinfo{author}{\bibfnamefont{S.}~\bibnamefont{Gutmann}},
  \bibinfo{author}{\bibfnamefont{J.}~\bibnamefont{Lapan}},
  \bibinfo{author}{\bibfnamefont{A.}~\bibnamefont{Lundgren}}, \bibnamefont{and}
  \bibinfo{author}{\bibfnamefont{D.}~\bibnamefont{Preda}},
  \bibinfo{journal}{Science} \textbf{\bibinfo{volume}{292}},
  \bibinfo{pages}{472} (\bibinfo{year}{2001}), \eprint{quant-ph/0104129}.

\bibitem[{\citenamefont{Garey and Johnson}(1979)}]{Garey}
\bibinfo{author}{\bibfnamefont{M.~G.} \bibnamefont{Garey}} \bibnamefont{and}
  \bibinfo{author}{\bibfnamefont{D.~S.} \bibnamefont{Johnson}},
  \emph{\bibinfo{title}{Computers and intractability: a guide to the theory of
  NP-completeness}} (\bibinfo{publisher}{W.H. Freeman and Company, San
  Francisco}, \bibinfo{year}{1979}).

\bibitem[{\citenamefont{Childs et~al.}(2002{\natexlab{a}})\citenamefont{Childs,
  Farhi, Goldstone, and Gutmann}}]{ChildsQIC}
\bibinfo{author}{\bibfnamefont{A.~M.} \bibnamefont{Childs}},
  \bibinfo{author}{\bibfnamefont{E.}~\bibnamefont{Farhi}},
  \bibinfo{author}{\bibfnamefont{J.}~\bibnamefont{Goldstone}},
  \bibnamefont{and} \bibinfo{author}{\bibfnamefont{S.}~\bibnamefont{Gutmann}},
  \bibinfo{journal}{Quant.\ Info.\ Comp.} \textbf{\bibinfo{volume}{2}},
  \bibinfo{pages}{181} (\bibinfo{year}{2002}{\natexlab{a}}),
  \eprint{quant-ph/0012104}.

\bibitem[{\citenamefont{Hogg}(2003)}]{Hogg}
\bibinfo{author}{\bibfnamefont{T.}~\bibnamefont{Hogg}}, \bibinfo{journal}{Phys.
  Rev. A} \textbf{\bibinfo{volume}{67}}, \bibinfo{pages}{022314}
  (\bibinfo{year}{2003}), \eprint{NB: quant-ph/0206059 (v2: Jan 2004) presents
  simulation data extended to 30 qubits}.

\bibitem[{\citenamefont{van Dam et~al.}(2001)\citenamefont{van Dam, Mosca, and
  Vazirani}}]{Vazirani}
\bibinfo{author}{\bibfnamefont{W.}~\bibnamefont{van Dam}},
  \bibinfo{author}{\bibfnamefont{M.}~\bibnamefont{Mosca}}, \bibnamefont{and}
  \bibinfo{author}{\bibfnamefont{U.}~\bibnamefont{Vazirani}},
  \bibinfo{journal}{Proc. 42nd Symp. FOCS} p. \bibinfo{pages}{279}
  (\bibinfo{year}{2001}), \eprint{quant-ph/0206003}.

\bibitem[{\citenamefont{Farhi et~al.}(2002)\citenamefont{Farhi, Goldstone, and
  Gutmann}}]{Farhi2}
\bibinfo{author}{\bibfnamefont{E.}~\bibnamefont{Farhi}},
  \bibinfo{author}{\bibfnamefont{J.}~\bibnamefont{Goldstone}},
  \bibnamefont{and} \bibinfo{author}{\bibfnamefont{S.}~\bibnamefont{Gutmann}}
  (\bibinfo{year}{2002}), \eprint{quant-ph/0201031}.

\bibitem[{\citenamefont{Childs et~al.}(2002{\natexlab{b}})\citenamefont{Childs,
  Farhi, and Preskill}}]{ChildsPRA}
\bibinfo{author}{\bibfnamefont{A.~M.} \bibnamefont{Childs}},
  \bibinfo{author}{\bibfnamefont{E.}~\bibnamefont{Farhi}}, \bibnamefont{and}
  \bibinfo{author}{\bibfnamefont{J.}~\bibnamefont{Preskill}},
  \bibinfo{journal}{Phys. Rev. A} \textbf{\bibinfo{volume}{65}},
  \bibinfo{pages}{012322} (\bibinfo{year}{2002}{\natexlab{b}}),
  \eprint{quant-ph/0108048}.

\bibitem[{\citenamefont{Roland and Cerf}(2003)}]{RolandCerf}
\bibinfo{author}{\bibfnamefont{J.}~\bibnamefont{Roland}} \bibnamefont{and}
  \bibinfo{author}{\bibfnamefont{N.~J.} \bibnamefont{Cerf}},
  \bibinfo{journal}{personal communication}  (\bibinfo{year}{2003}).

\bibitem[{\citenamefont{Mooij et~al.}(1999)\citenamefont{Mooij, Orlando,
  Levitov, Tian, van~der Wal, and Lloyd}}]{Mooij}
\bibinfo{author}{\bibfnamefont{J.~E.} \bibnamefont{Mooij}},
  \bibinfo{author}{\bibfnamefont{T.~P.} \bibnamefont{Orlando}},
  \bibinfo{author}{\bibfnamefont{L.}~\bibnamefont{Levitov}},
  \bibinfo{author}{\bibfnamefont{L.}~\bibnamefont{Tian}},
  \bibinfo{author}{\bibfnamefont{C.~H.} \bibnamefont{van~der Wal}},
  \bibnamefont{and} \bibinfo{author}{\bibfnamefont{S.}~\bibnamefont{Lloyd}},
  \bibinfo{journal}{Science} \textbf{\bibinfo{volume}{285}},
  \bibinfo{pages}{1036} (\bibinfo{year}{1999}).

\bibitem[{\citenamefont{Orlando et~al.}(1999)\citenamefont{Orlando, Mooij,
  Tian, van~der Wal, Levitov, Lloyd, and Mazo}}]{Orlando}
\bibinfo{author}{\bibfnamefont{T.~P.} \bibnamefont{Orlando}},
  \bibinfo{author}{\bibfnamefont{J.~E.} \bibnamefont{Mooij}},
  \bibinfo{author}{\bibfnamefont{L.}~\bibnamefont{Tian}},
  \bibinfo{author}{\bibfnamefont{C.~H.} \bibnamefont{van~der Wal}},
  \bibinfo{author}{\bibfnamefont{L.~S.} \bibnamefont{Levitov}},
  \bibinfo{author}{\bibfnamefont{S.}~\bibnamefont{Lloyd}}, \bibnamefont{and}
  \bibinfo{author}{\bibfnamefont{J.~J.} \bibnamefont{Mazo}},
  \bibinfo{journal}{Phys. Rev. B} \textbf{\bibinfo{volume}{60}},
  \bibinfo{pages}{15398} (\bibinfo{year}{1999}).

\bibitem[{\citenamefont{Barahona}(1982)}]{Barahona}
\bibinfo{author}{\bibfnamefont{F.}~\bibnamefont{Barahona}},
  \bibinfo{journal}{J. Phys. A} \textbf{\bibinfo{volume}{15}},
  \bibinfo{pages}{3241} (\bibinfo{year}{1982}).

\bibitem[{\citenamefont{Garey et~al.}(1976)\citenamefont{Garey, Johnson, and
  Stockmeyer}}]{Garey2}
\bibinfo{author}{\bibfnamefont{M.~G.} \bibnamefont{Garey}},
  \bibinfo{author}{\bibfnamefont{D.~S.} \bibnamefont{Johnson}},
  \bibnamefont{and}
  \bibinfo{author}{\bibfnamefont{L.}~\bibnamefont{Stockmeyer}},
  \bibinfo{journal}{Theo. Comp. Sci.} \textbf{\bibinfo{volume}{1}},
  \bibinfo{pages}{237} (\bibinfo{year}{1976}).

\bibitem[{\citenamefont{Baker}(1994)}]{Baker}
\bibinfo{author}{\bibfnamefont{B.~S.} \bibnamefont{Baker}},
  \bibinfo{journal}{J. of the ACM} \textbf{\bibinfo{volume}{41}},
  \bibinfo{pages}{153} (\bibinfo{year}{1994}).

\bibitem[{\citenamefont{Storcz and Wilhelm}(2003)}]{Storcz}
\bibinfo{author}{\bibfnamefont{M.~J.} \bibnamefont{Storcz}} \bibnamefont{and}
  \bibinfo{author}{\bibfnamefont{F.~K.} \bibnamefont{Wilhelm}},
  \bibinfo{journal}{Appl. Phys. Lett.} \textbf{\bibinfo{volume}{83}},
  \bibinfo{pages}{2387} (\bibinfo{year}{2003}).

\end{thebibliography}

\end{document}